\documentclass[preprint,12pt]{elsarticle}
\usepackage{amssymb}
\usepackage{graphicx}
\usepackage{natbib}
\usepackage{amsmath}

\newcommand{\xvect}{\mbox{\bf x}}

\usepackage{lineno}

\journal{Chaos, Solitons and Fractals}

\begin{document}

\begin{frontmatter}
\title{Energy analysis of bursting Hindmarsh-Rose neurons with time-delayed coupling.}
\author{A. Moujahid and F. Vadillo}
\address{Department of Mathematics, University of the
Basque Country, UPV/EHU, Spain}

\date{\today}

\begin{abstract}
Mathematical modeling is an important tool to study the role of delay in neural systems and to evaluate its effects on the signaling activity of coupled neurons. Models for delayed neurons are often used to represent the dynamics of real neurons, but rarely to assess the energy required to maintain these dynamics. In this work, we address these questions from an energy perspective by considering a pair of Hindmarsh-Rose burst neurons coupled by reciprocal time-delayed coupling with electrical and chemical synapses. We examine the average energy consumption required to maintain cooperative behavior and quantify the contribution of synapses to total energy consumption. We show that unlike electrical coupling, where the time delay appears to reduce the instantaneous average relative weight of the synaptic contribution, in chemical coupling this average synaptic contribution appears to be much higher in delayed coupling than in instantaneous coupling, except at certain values of coupling strength where the instantaneous synaptic contribution is more important.
\end{abstract}


\end{frontmatter}

\maketitle

\section{Introduction}
\medskip

The oscillatory pattern that emerges when a coupling is introduced between two neurons depends on their intrinsic dynamics as well as on the nature of the coupling. Bursting is a type of oscillatory regime that occurs when the activity of neurons alternates between a quiescent state and rapid, repetitive spikes \cite{WANG1993,Izhikevich2000}. Much research has been done on the mechanisms that give rise to such bursting \cite{Izhikevich2000,Izhikevich2003,Pinto2000}, and has been reported that bursting synchronization can be related to a number of pathologies such as Parkinson’s disease, epilepsy and essential tremors \cite{Benito2019,Adriane2021}.

Unlike coupled spike neurons, whose synchronous dynamics are relatively simple, interacting burst neurons can exhibit a wide variety of possible forms of synchrony, including single-spike synchrony, burst synchrony, and full synchrony \cite{Elson1998,Vreeswijk2001,Dhamala2004}. Typically, burst synchrony occurs with low coupling strength, whereas full synchrony, which includes both burst and spike synchrony, requires stronger coupling.

The synchronization of bursts in instantaneously electrically coupled HR neurons has been largely studied in literature \cite{Dhamala2004,Erichsen2006,Torrealdea2009,Moujahid2011}, and it was shown that sufficiently strong coupling leads to full synchrony. Unlike electrical coupling, instantaneously chemical coupling seems not to be favorable to produce bursts synchronization for any value of coupling strength. However, time-delayed coupling, has proven to promote exactly synchronous bursting dynamics for all values of the coupling that imply bursting, provided that the time delay is in the appropriate range \cite{Li2007,Liang2009}.

Delays are indeed found in many biological, physical, and engineering systems. In physiological systems, time delays occur almost everywhere, which is a natural consequence of the limited speed with which physiological or chemical processes are transmitted from one place to another \cite{BATZEL2011}. In the brain, for example, the transmission of information is delayed due to the limited speed of signal propagation in the nervous system. It has been shown that the time delay has a significant influence not only on the dynamics of the signal activity of neurons, but also on synchronization phenomena and the transition between different synchronization states, and that it is responsible for several interesting phenomena in coupled dynamical systems \cite{BATZEL2011,Campbell2007}.

The study of synchronization transitions in small-world neural networks modeled locally by the Rulkov map has shown that phenomena such as zigzag fronts of excitations, anti-phase synchronization, or in-phase synchronization occur as a function of the magnitude of the time delay \cite{Wang_et_al_2008}. In addition, increasing delay has been suggested to be responsible for the intermittent occurrence of regions of irregular and regular excitation fronts in scale-free neural networks \cite{Wang_Perc_2009} and also for the wavelike synchronization of bursting oscillations in the presence of attractive and repulsive coupling \cite{Wang_et_al_2011}.
Other authors have studied the complex interplay between time delay, ephaptic coupling, and synchrony in multilayer neural networks and have shown that weak coupling can improve synchrony in the presence of partial time delay in chemical communication \cite{SHAFIEI2020}.

In this paper, we address the effects of time delay on the synchronization of coupled neurons from an energetic perspective. Indeed, the physical concept of energy provides a solid approach to characterize not only a physical system itself, but also the interactions between the parts of this system as well as its interactions with other systems.

We study the energy requirements involved in the synchronization of two coupled Hindmarsh-Roses neurons with symmetric time-delayed coupling. We consider both electrical and chemical coupling. Since most of the brain's energy is used for synaptic transmission, we focus on the relative weight of the synapse's contribution to the total energy income for different values of synaptic coupling strength and time delays \cite{Buric2008}.

The remainder of the paper is organized as follows. Section \ref{sec2} describes the nature of the delayed coupling mechanisms. Section \ref{sec3} presents the energy approach used to calculate the relative weight of the synapse's contribution to the total energy received by a neuron through the membrane. We also give a formula for the average net energy that characterizes the coupled system. Section \ref{sec3} describes the Hindmarsh-Rose neuron model and the energy function associated with its dynamics. The results obtained are discussed in section \ref{sec5}. Finally, section \ref{sec6} concludes the paper.

\section{Delayed Coupled Neurons}
\label{sec2}
In this section, we introduce the class of electrically and chemically delayed coupling, which is commonly used in the literature. Let us consider two oscillating neurons $\dot{\xvect}_i=f_i(\xvect_i)$ and $\dot{\xvect}_j=f_j(\xvect_j)$.

For chemical coupling, the neurons are coupled according to the so-called fast threshold modulation scheme with explicit time delay \cite{Somers1993}:
\begin{equation}
\begin{split}
\dot{\xvect}_i=f_i(\xvect_i)+g_c(V_s-\xvect_i) \Gamma(\xvect_j^{\tau}),\\
\end{split}
\label{equ01}
\end{equation}
where the indices $i,j=1,2 (i\neq j)$ refer to the postsynaptic and presynaptic neurons, respectively.  $\xvect_i,\xvect_j\in\Re^n$ denotes the state of the coupled neurons, $f_i,f_j\in \Re^n\to \Re^n$ are smooth functions, and $g_c$ is the synaptic coupling strength between neuron $i$ at time $t$ and neuron $j$ at an earlier time $t-\tau$. $\Gamma$ is the sigmoid function given by,
\[
\begin{split}
\Gamma(x_j^{\tau})=\frac{1}{1+exp(-k(x_j^{\tau}-\theta_s))},\\
\end{split}
\label{equ01a}
\]
and $x_j^{\tau}=x_j(t-\tau)$.

The parameters of the chemical synapses are assumed to be symmetric, i.e, no indices in the parameter $g_c$. The implemented sigmoid threshold function can be modified to account for both soft-threshold-like behavior, similar to graded synaptic transmission, and hard-threshold-like behavior, related to the fast threshold modulation model.

This coupling scheme allows either a hard or a gradual threshold behavior depending on the synaptic gain parameter $k$. The reversal potential $V_s$ determines the character of the coupling. When $V_s$ is smaller than the postsynaptic membrane potential $x_i$, the synaptic current has a hyperpolarizing effect and leads to an inhibitory synapse. A larger reversal potential, on the other hand, has a depolarizing effect and makes the synapse excitatory.
In this work, we fixed these parameters to the values given in reference \cite{Buric2008}, which correspond to the models of fast threshold modulation commonly observed for most chemical synapses in the brain, i.e., $\theta_s=-0.25$, $V_s=2$, and $k=10$.

For electrical coupling, the two neurons are coupled using the delayed linear coupling as follow:
\begin{equation}
\begin{split}
\dot{\xvect}_i=f_i(\xvect_i)+g_c\left(\xvect_j(t-{\tau})-\xvect_i\right).\\
\end{split}
\label{equ02}
\end{equation}

\section{Energy contribution from delayed synapses}
\label{sec3}
In this section, we apply the energy approach we developed in previous work \cite{Torrealdea2009,Moujahid2011,Sarasola2004} to study the energy contribution of these chemical synapses, and how they are affected by the synaptic coupling strength ($g_c$) and the time delay ($\tau$).

We assume that the dynamical system $\dot{\xvect}=f(\xvect)$, has an energy function, $H({\xvect})$, characterizing its dynamics.
This energy is compatible with the generalized Hamiltonian for the conservative system  $\dot{\xvect}=f_c(\xvect)$, where $f_c$ is the divergence-free vector field associated with $f(\xvect)$. The variation of this energy is mainly due to the dissipative component of the velocity vector field $f(\xvect)$ according to the equation \cite{Sarasola2004},
\begin{equation}
\begin{split}
\dot{H}=\nabla H^T f_d(\xvect),\\
\end{split}
\label{equ03}
\end{equation}
where $f_d$ refers to the curl-free vector carrying the total divergence of the vector field $f(\xvect)$. This energy function is used to evaluate the energy demand of a neuron's signaling activity, both when it is isolated and when it is connected to other neurons via chemical synapses. It forms the basis for all computational results presented in this paper.

For an isolated system, the long-term average of energy derivative given by Eq. \ref{equ03} is zero. The average involves a global energy balance between the components responsible for energy consumption and the components that supply energy. According to this scheme, the metabolic energy that must be supplied to the neuron to maintain its activity is equal to the long-term average of only the positive component of the energy derivative $\dot{H}$, which we denote by $\langle\dot{H}^+\rangle_m$. In contrast, the neuron's energy loss through its membrane is given by the long-term average of the negative component of $\dot{H}$, which we denote by $\langle\dot{H}^-\rangle_m$. The subscript $m$ refers to the membrane.

When two neurons force each other, their respective dynamics change, although they still remain within an attractive region of phase space. When the coupling is electrical, the long-term average net energy of the augmented system given by $\left(\dot{\xvect}_i=f(\xvect_i)+g_c(\xvect_j(t-\tau)-\xvect_i)\right)$ tends to zero \cite{Sarasola2004}, mainly due to the nature of the coupling term, which tends to zero when the two neurons reach a synchronized state. However, when a chemical synapse is involved, the extended system is given by,
$\left(\dot{\xvect}_i=f(\xvect_i)+g_c(Vs-\xvect_i) \Gamma(\xvect_j^{\tau})\right)$, where a synchronized state is no longer characterized by a zero coupling term. To understand how this kind of coupling affects the energy balance of coupled neurons, we proceed as follows.

According to our energy formalism, the net average energy of the augmented system is given by,
\begin{equation}
\begin{split}
\left<\nabla H^T f_d(\xvect_i)\right> + \left<\nabla H^T g_c(Vs-\xvect_i) \Gamma(\xvect_j^{\tau}) \right>\\
\end{split}
\label{equ04}
\end{equation}
where $\nabla H^T$ denotes the transpose gradient of the energy function $H({\xvect})$. The brackets refer to the averaging over the attractor.

The first term of Eq. \ref{equ04}, which corresponds to the average energy change of a neuron through its membrane, can be decomposed into the energy that a neuron takes up across its membrane, $\langle\dot{H}^+\rangle_m^i$, and its energy output across the membrane,$\langle\dot{H}^-\rangle_m^i$.  When there is no synaptic transmission (i.e., when $g_c=0$), these two terms are perfectly balanced, and Eq.  \ref{equ04} reduces to zero.

When the neurons are coupled, the average energy change of a neuron through its membrane, $\langle\dot{H}\rangle_m^i$, should be  balanced by its average energy through the synapse, $\langle\dot{H}\rangle_s^i$ (second term of Eq.\ref{equ04}).

Thus, the energy balance given by Eq. \Ref{equ04}, can be formulated as follows:
\begin{equation}
\begin{split}
\langle\dot{H}^+\rangle_m^i+\langle\dot{H}^-\rangle_m^i+\langle\dot{H}\rangle_s^i
\end{split}
\label{equ05}
\end{equation}

The relative weight of the contribution of the synapse, $S_w$,  to the total energy income through the membrane can be estimated as follows,
\begin{equation}
\begin{split}
S_w=\frac{\langle\dot{H}\rangle_s^i}{\langle\dot{H}^+\rangle_m^i}
\end{split}
\label{equ06}
\end{equation}
The equations \ref{equ05} and \ref{equ06} form the basis of computational results presented in this paper.


\section{The Hindmarsh-Rose neuron}
\label{sec4}
The four-dimensional extension of the original Hindmarsh-Rose model \cite{Hindmarsh1984,Rose1985} is characterized by having much larger regions in parameter space where chaotic behavior occurs. This model exhibits dynamic behavior that is similar in several aspects to the properties of real biological neurons, and is described by the following equations of motion:
\begin{equation}
\begin{cases}
\dot{x}=& ay+bx^2-cx^3-dz+\xi I, \\
\dot{y}=&  e-fx^2-y-gw, \\
\dot{z}=&  m(-z+s(x+h)), \\
\dot{w}=&  n(-kw+r(y+1)),
\end{cases}
\label{equ07}
\end{equation}
where $x$ represents the membrane voltage, while $y$ and $z$ describe some fast and slow gating variables for ionic currents, respectively. The variable $w$ represents an even slower dynamic variable and was introduced because a slow process such as calcium exchange between intracellular stores and the cytoplasm was found to be required to fully reproduce the observed chaotic oscillations of isolated neurons from the stomatogastric ganglion of the California spiny lobster Panulirus interruptus \cite{Pinto2000}.

\begin{table}[htb]\label{table1}
    \centering
    \begin{tabular}{c|l|c}\hline
        Parameter&Value&Unit\\\hline
        $a$&1.0&-\\
        $b$&3.0 &$(mV)^{-1}$\\
        $c$& 1 &$(mV)^{-2}$\\
        $d$& 0.99&$M\Omega$\\
        $\xi $&1&$ M\Omega$\\
        $e$&1.01&$mV$\\
        $f$&5.0128&$ (mV)^{-1}$\\
        $g$&0.0278&$M\Omega$\\
        $m$&0.00215&-\\
        $s$& 3.966&$ \mu S$\\
        $h$& 1.605&$mV$\\
        $n$&0.0009&-\\
        $k$&0.9573&-\\
        $r$& 3.0&$ \mu S$\\
        $l$& 1.619&$mV$\\\hline
    \end{tabular}
    \caption{The parameters of the Hindmarsh-Rose neuron model as reported in \cite{Pinto2000}.}
    \label{tab:param}
\end{table}

For the numerical results of this paper we fix the parameters to the one given in Table \ref{tab:param}

\begin{figure}[hbt!]
  \centering
  \includegraphics[width=.95\textwidth]{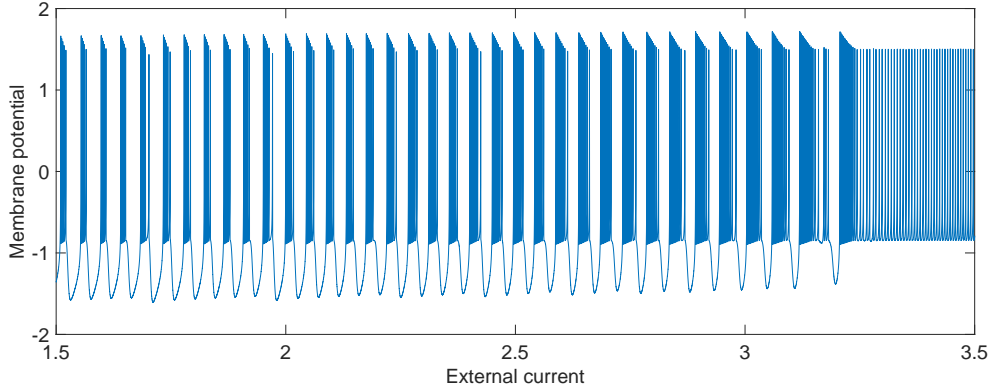}
  \caption{Membrane voltage x(t) as a function of external stimulus $I$ of the Hindmarsh-Rose model given by Eq. \ref{equ07}, simulated with the parameters given in Table 1. To capture the different oscillatory behaviors of the model, we considered an instantaneous linear increase of I from 1.5 nA to 3.5 nA. }\label{membranePotentialVersusI}
\end{figure}

The parameter $I$ in Eq.(\ref{equ07}) represents the external current input and is a bifurcation parameter that characterises the model dynamics. Figure \ref{membranePotentialVersusI} shows the instantaneous time serie of the membrane voltage $x(t)$ for a linearly increase of the the external current from 0 to 4 nA. As one can see, spiking activity smoothly alternates between different states such as rest, bursting, and tonic spiking.

In a previous paper we assigned an energy function to the Hindmarsh-Rose model given by equation (\ref{equ05}). The procedure for finding this energy was described in detail in \cite{Torrealdea2009}. This energy function $H(\mathbf x)$ is given by
\begin{equation}
\begin{array}{ll}
 H=\frac{p}{a}(\frac{2}{3}fx^3+\frac{msd-gnr}{a}x^2+ay^2)\\
+\frac{p}{a}(\frac{d}{ams}(msd-gnr)z^2-2dyz+2gxw)
\label{equ08}
\end{array}
\end{equation}
Since each adding term in equation (\ref{equ08}) has the dimension of the quadratic voltage, and subject to the condition that the parameter $p$ has the dimension of the conductance, the function $H$ coincides dimensionally with a physical energy. In this paper, we fix this parameter to the arbitrary value $p=-1 S$. We have assumed a negative sign for this parameter so that the result of the model is consistent with the usual assumption of an energy demand associated with the repolarization period of the membrane potential and also with its refractory period.

The energy variation across the membrane is given by,
\begin{equation}
\begin{split}
 \dot{H}&=\frac{2p}{a}\left[ (fx^2+\frac{msd-gnr}{a}x+gw)(bx^2-cx^3+\xi I)\right]\\
 &+\frac{2p}{a}\left[(ay-dz)(e-y)\right]\\
 &+\frac{2p}{a}\left[(\frac{d}{ams}(msd-gnr)z-dy)(msh-mz_i)\right]\\
 &+\frac{2p}{a}\left[gx(nrl-nkw)\right]
\end{split}
\label{equ09}
\end{equation}

This energy variation is the sum of the energy input ($\dot{H}^+_m$) through the membrane plus the energy dissipated through the membrane ($\dot{H}^-_m$) . These two components were calculated using the positive and negative parts of $\dot{H}$, respectively.

On the other hand, if two neurons are coupled according to Eq. \ref{equ12}, the energy change at the synapse is given as follows,
\begin{equation}
\begin{split}
 \dot{H}_s^i&=\frac{2p}{a}\left[ (fx_i^2+\frac{msd-gnr}{a}x_i+gw_i)g_c(Vs-x_i)\Gamma(x_j^{\tau})\right]
\end{split}
\label{equ10}
\end{equation}
When the neurons are coupled via electrical synapses (see Eq. \ref{equ13}) , this energy is given by,
\begin{equation}
\begin{split}
 \dot{H}_s^i&=\frac{2p}{a}\left[ (fx_i^2+\frac{msd-gnr}{a}x_i+gw_i)g_c(x_j(t-\tau)-x_i)\right]
\end{split}
\label{equ11}
\end{equation}

\begin{figure}[hbt!]
  \centering
  \includegraphics[width=.95\textwidth]{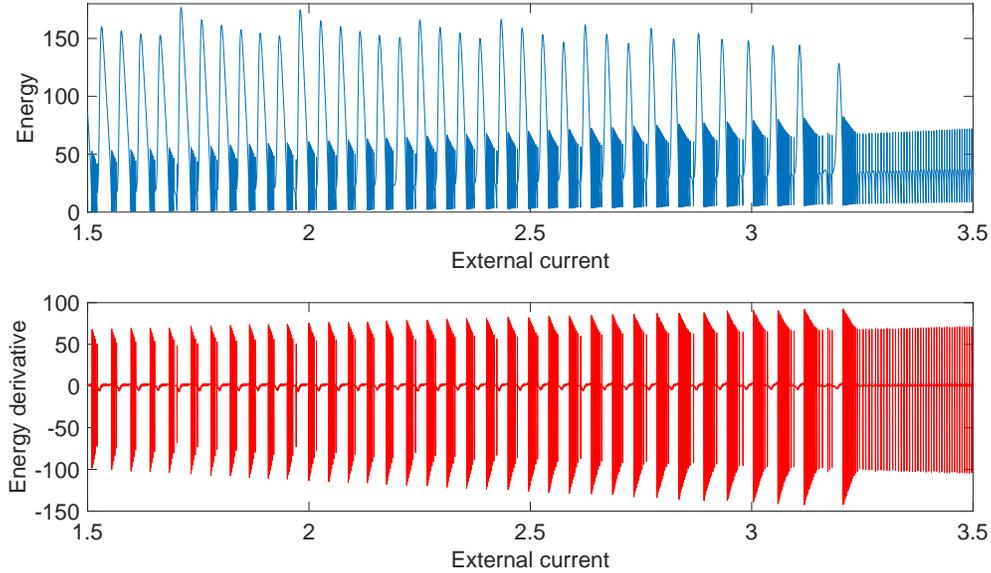}
  \caption{Instantaneous energy $H(t)$ according to Eq. 8 (top) and energy derivative $\dot{H}(t)$ according to Eq. 9 (bottom) as a function of external current in nA. The energy values are negative, but have been presented with a positive sign to better illustrate the energy required to generate a spike. The simulation setting is the same as in Figure \ref{membranePotentialVersusI}.}\label{energiesVersusI}
\end{figure}

Figure \ref{energiesVersusI} shows the oscillatory pattern of energy and energy dissipation that characterizes the different  regimes that occur when the external current I is linearly increased from 0 to 4 nA. During the refractory period between two successive spikes, the energy dissipation remains slightly positive, indicating that energy continues to be required until the onset of the action potential. During the rising phase of the action potential, the energy dissipation drops rapidly to negative and eventually to positive values, contributing to the repolarization of the membrane potential during the falling phase.

\section{Numerical results}
\label{sec5}
In this section, we consider two Hindmarsh-Rose neurons coupled with delayed synapses according to the schemes given by Eqs. \ref{equ01} and \ref{equ02}. That is, we consider the following system for chemical coupling:
\begin{equation}
\begin{cases}
\dot{x_i}=& ay_i+bx^2-cx^3-dz_i+\xi I+g_c(Vs-x_i)\Gamma(x_j^{\tau}), \\
\dot{y_i}=&  e-fx_i^2-y-gw_i, \\
\dot{z_i}=&  m(-z_i+s(x_i+h)), \\
\dot{w_i}=&  n(-kw_i+r(y_i+1)),
\end{cases}
\label{equ12}
\end{equation}
where $\Gamma(x_j^{\tau})=\frac{1}{1+exp(-k(x_j(t-\tau)-\theta_s))}$. And, for electrical coupling, we have:
\begin{equation}
\begin{cases}
\dot{x_i}=& ay_i+bx^2-cx^3-dz_i+\xi I+g_c\left(x_j(t-\tau)-x_i\right), \\
\dot{y_i}=&  e-fx_i^2-y-gw_i, \\
\dot{z_i}=&  m(-z_i+s(x_i+h)), \\
\dot{w_i}=&  n(-kw_i+r(y_i+1)),
\end{cases}
\label{equ13}
\end{equation}

Note that the coupling only affects the membrane voltage variables.

For the numerical results, we adopted the parameter values given in Table \ref{tab:param} and considered symmetric coupling. We assume that the two neurons are confined in the chaotic regime corresponding to an external current $I=3.024$. All energy-based measurements have been averaged  over 20,000 different spike trains of 25ms length.

To understand the energy fluxes across the membrane involved in both electrical and chemical synapses, we calculated the average energy balance per unit time, according to Eq. \ref{equ05} required to maintain the organised behaviour of the coupled neurons, and the average relative weight of the synaptic contribution according to Eq. \ref{equ06}.

\subsection{Instantaneous coupling ($\tau=0$)}
\begin{figure}[hbt!]
  \includegraphics[width=.95\textwidth]{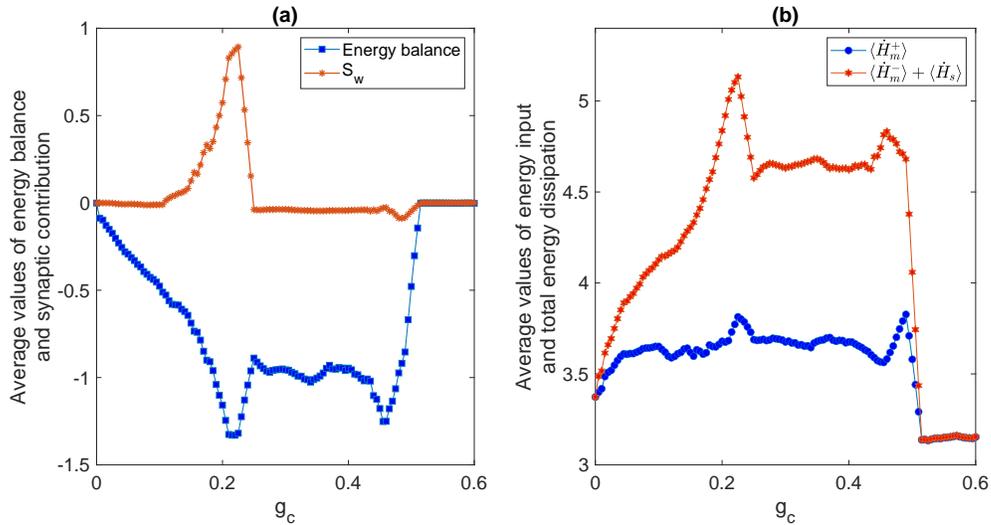}
  \caption{{\bf Instantaneous electrical coupling.} {\bf (a)} the average energy balance per unit time  according to Eq. \ref{equ05} and the average relative weights of synaptic contribution ($S_w$) according to Eq. \ref{equ06} at different values of synaptic coupling strength $g_c$. {\bf (b)} the average total energy input through the membrane ($\langle \dot{H}_m^+\rangle$) and the average dissipation through the membrane plus the net flow of energy in the synapse ($\langle \dot{H}_m^-\rangle+\langle\dot{H}_s \rangle$). The coupled neurons are stimulated by an external current $I=3.024$, corresponding to a chaotic regime.}\label{figure3}
\end{figure}

\begin{figure}[hbt!]
  \includegraphics[width=.95\textwidth]{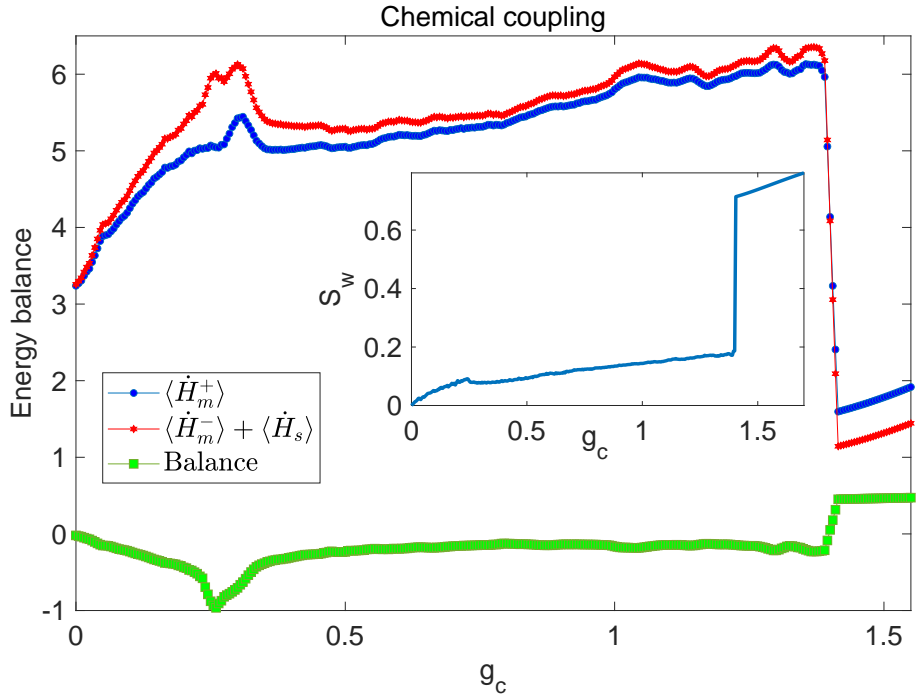}
  \caption{{\bf Instantaneous chemical coupling} The average energy balance per unit time  according to Eq. \ref{equ05} (green square markers) and the average relative weights of synaptic contribution, $S_w$, according to Eq. \ref{equ06} (see inset) at different values of synaptic coupling strength $g_c$.
  We also plotted the average total energy input through the membrane ($\langle \dot{H}_m^+\rangle$, blue circle markers) and the average dissipation through the membrane plus the net flow of energy in the synapse ($\langle \dot{H}_m^-\rangle+\langle\dot{H}_s \rangle$, red hexagram markers).}\label{figure4}
\end{figure}

In this subsection, we perform a quantitative comparison of energy balance and synaptic contribution in instantaneous chemical and electrical couplings. The results are shown in figures \ref{figure3} and \ref{figure4}.

\begin{figure}[hbt!]
  \centering
      \includegraphics[width=.95\textwidth]{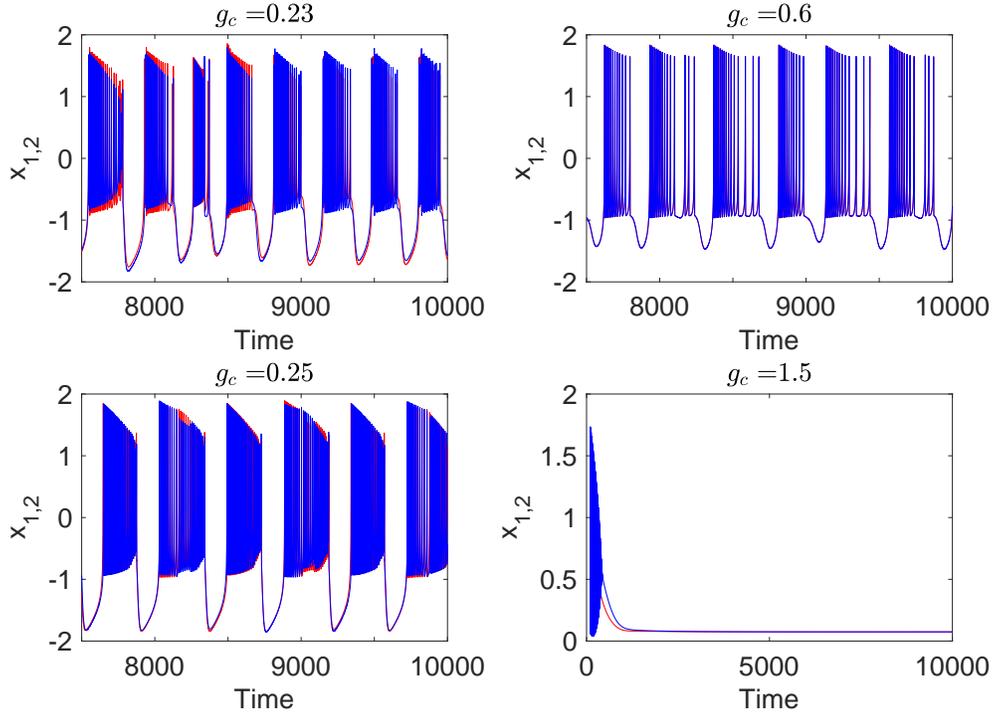}
  \caption{{\bf Instantaneous coupling.} Long-term trajectories of coupled neurons at different values of synaptic coupling strength $g_c$. {\bf Top panels:} electrical coupling.    {\bf Bottom panels:} chemical coupling.}\label{figure5}
\end{figure}

In the case of electrical coupling (see Figure \ref{figure3}(a)), we see that initially, when $g_c=0$, the energy balance is zero, meaning that the average energy dissipated by the membrane is balanced by the average energy absorbed by the membrane. Once coupling is established, the coupled neurons are forced to oscillate in manifolds where the long-term average of their energy balance is no longer zero.
In the region of synaptic coupling strength $g_c\in[0,0.5]$, which corresponds to a region of transition to the synchronized regime, the energy balance is even negative, meaning that the energy input is less than the energy output.
As can be seen in panel (b), the average total energy input (blue curve with circle marker) is less than the average energy dissipation through the membrane plus the net energy flux in the synapse (red curve with hexagram marker).
This means that the activity of the neurons is more energetically demanding and additional energy flux is required for collective behavior.
On the other hand, the synapse makes an important contribution around $g_c=0.23$, accounting for up to 90\% of the total energy supplied to the neuron across its membrane (see panel (a) red curve with star marker). It appears that at this strength of synaptic coupling, neurons begin to synchronize their bursts (see Figure \ref{figure5}, top, left). Synchronization of burst neurons is indeed a phenomenon on multiple time scales, where there are two distinct transitions to synchronized states, one associated with bursts and the other with spikes \cite{Dhamala2004}.
After that, the energy balance is restored at $g_c=.5$ when the neurons synchronize completely (see Figure \ref{figure5}, Top, right).

For chemical coupling (see Figure \ref{figure4}), the average energy balance shows a similar pattern to that of electrical coupling.
In the region where the neurons are not yet synchronized, the average energy balance quickly drops to negative values as soon as the coupling is established. After reaching a minimum at $g_c=0.25$, the balance is partially restored, with values approaching zero. At this synaptic coupling strength, the neurons experience a weak type of synchrony, but the spikes within the bursts are not synchronized (see Figure \ref{figure5}, bottom left). This state is maintained until $g_c=1.44$, where dissipation through the membrane is slightly greater than its total average income. A further increase in synaptic coupling strength causes the coupled neurons to reach a stationary synchronized state (see Figure \ref{figure5}, bottom right), corresponding to a positive energy balance. This fact is a consequence of the different oscillatory behavior of the neuron when coupled and seems to depend on the type of coupling used for synchronization.

\subsection{Delayed coupling ($\tau>0$)}

As mentioned in the introduction, synaptic transmission time is particularly important in chemical-type synapses as opposed to electrical synapses. It is well known that the explicit time delay in modeling a synaptic connection of physically appropriate duration can have profound effects on the dynamics of the coupled neurons. Indeed, the time delay can enhance burst synchronization between coupled neurons. In this subsection, we examine the average energy balance under time-delayed conditions and how energy demands can be reduced by an appropriate choice of synaptic coupling strength and time delay.

\begin{figure}[hbt!]
  \centering
      \includegraphics[width=.95\textwidth]{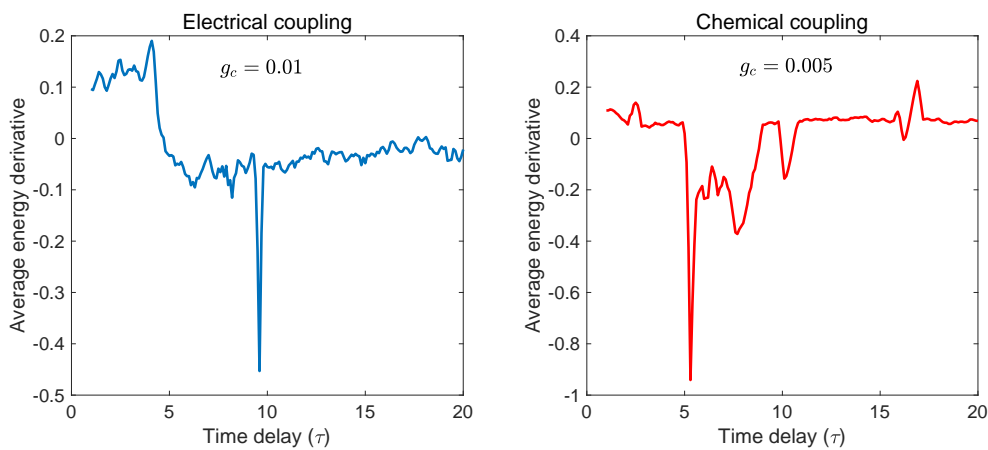}
  \caption{{\bf Average energy derivative} according to Eq. \ref{equ09} as a function of the time delay $\tau$ for very small synaptic coupling strengths. {\bf Left:} electrical coupling with $g_c=0.01$. {\bf Right:} chemical coupling with $g_c=0.005$   .} \label{figure6}
\end{figure}


\begin{figure}[hbt!]
  \centering
      \includegraphics[width=1\textwidth]{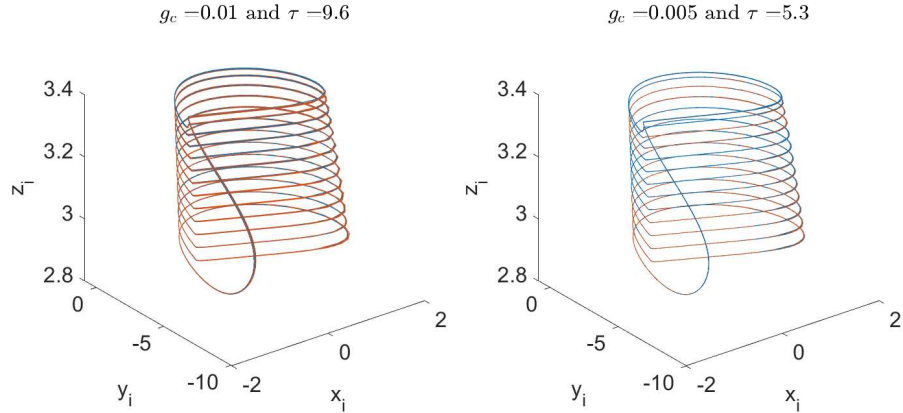}
  \caption{Long-term trajectories of coupled neurons at the values of synaptic coupling strength $g_c$ and time delay $\tau$ reported in Figure \ref{figure6} . {\bf Left: } electrical coupling.    {\bf Right:} chemical coupling} \label{figure7}
\end{figure}

First, we analysed the average energy derivative, that is the long term average of Eq. \ref{equ09}, as a function of time delay for weak synaptic coupling strength. The results are shown in Figure \ref{figure6}. As one can see, there is a range of time delay that is energetically more favorable with low net dissipation of energy. For electrical coupling (Fig.\ref{figure6}, panel (a)), the synaptic coupling strength is set to $g_c=0.01$. In this case, the minimum average energy derivative occurs at $\tau=9.6$. For chemical coupling (Fig.\ref{figure6}, panel (b)), the synaptic coupling strength is set to small value $g_c=0.005$. The average energy derivative shows a global minimum at $\tau=5.3$, but we can also identify other regions where the activity of the coupled neurons is less energetically demanding. The lon-term trajectories of the coupled neurons corresponding to the configurations $(g_c=0.01, \tau=9.6)$ and $(g_c=0.005, \tau=5.3)$ are reported in Figure \ref{figure7}. As it can be seen, at these values of $g_c$ and $\tau$ the coupled neurons show synchronized regime at low values of coupling strength.

\begin{figure}[hbt!]
  \begin{tabular}{c}
       \includegraphics[width=.95\textwidth]{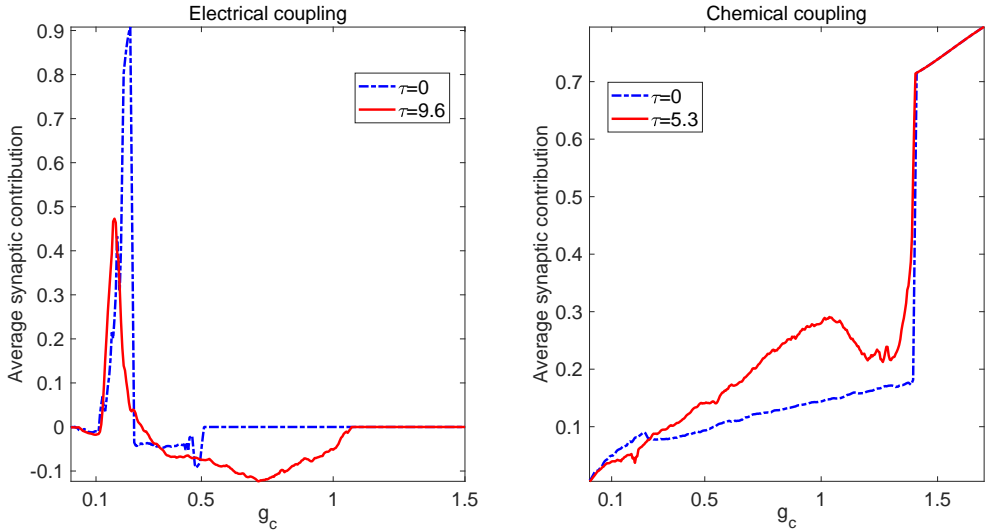}
  \end{tabular}
  \caption{{\bf Average synaptic contribution} for both electrical ({\bf Left}) and chemical ({\bf Right}) coupling as a function of synaptic strength at the values of $\tau$ reported in Figure \ref{figure6}.}\label{figure8}
  \label{figure8}
\end{figure}

\begin{figure}[hbt!]
  \begin{tabular}{c}
       \includegraphics[width=.95\textwidth]{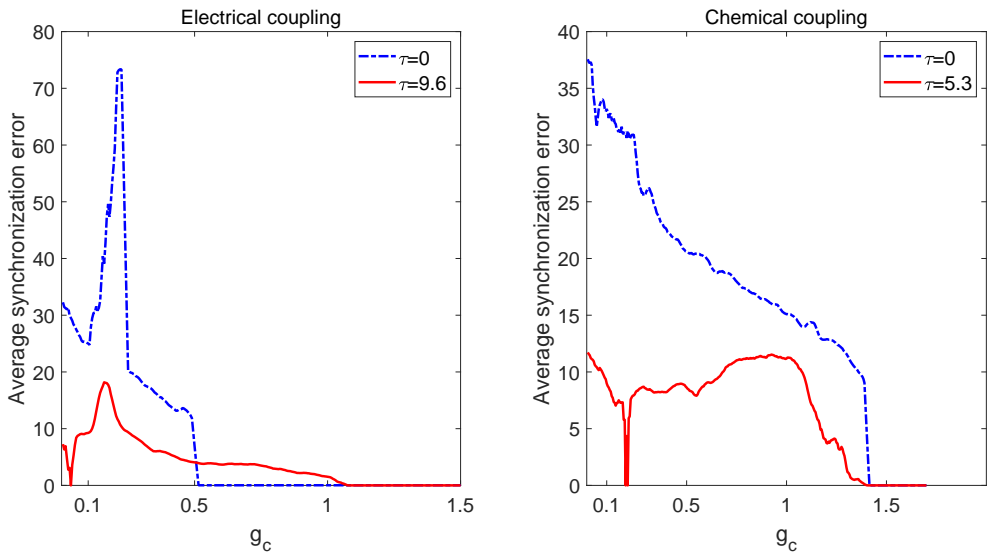}
  \end{tabular}
  \caption{{\bf Average synchronization error} between the coupled neurons at different values of synaptic coupling strength $g_c$. Synchronization error was measured as the norm of the error vector $e = ||x_i - x_j||$. {\bf Left:} electrical coupling.  {\bf Right:} chemical coupling.}\label{figure9}
\end{figure}

Figure \ref{figure8} shows a comparison of the average synaptic contribution between instantaneous and delayed coupling by electrical (left panel) and chemical (right panel) synapses.

For instantaneous electrical coupling (see the blue dashed curve in panel (a)), there is no significant synaptic contribution except in the range of values of synaptic strength $0.15 \leq g_c\leq 0.25$, where the maximum synaptic contribution occurs. In this range, the synapse contribution is up to 90\% of the total energy. The introduction of a time delay, on the other hand, appears to reduce this maximum relative weight of the synaptic contribution (see red solid curve in panel (a)), which falls to a value as low as 50\%. However, the time delay causes the threshold for full synchronization to increase 2-fold. In fact, with delayed electrical coupling, full synchronization is achieved at values of $g_c\approx 1$, whereas with instantaneous coupling, neurons reach full synchronization at $g_c=0.5$.

Unlike electrical coupling, the average synaptic contribution to total energy income shows an increasing pattern with synaptic weight and is much higher for delayed coupling than for instantaneous coupling, except for values of $g_c < 0.26$, where instantaneous synaptic contribution is more important (see \ref{figure8}- panel (b)).

As for the average synchronization error as a function of synaptic weight (see figure \ref{figure9}), we see that for both electrical and chemical coupling, the average error is much lower for delayed coupling than for instantaneous coupling. Furthermore, there are regions in $(g_c,\tau)$-space where the coupled neurons could achieve synchronous behaviors with zero error at very low values of the coupling strength $g_c$ (see red solid curves in panels (a) and (b)). These synchronous patterns are not observed in instantaneous coupling because there is a threshold above which synchronization is achieved.
For chemical coupling (see panel (b)), the threshold for full synchronization appears to be the same for both instantaneous and delayed synapses.

\section{Conclusions}
\label{sec6}
We analyzed both instantaneous and delayed electrical and chemical coupling modes from an energetic point of view. To do so, we used the energy approach developed in \cite{Sarasola2004} and focused our analysis on the average contribution of synapses to the total energy received by a neuron across the membrane. We also performed a quantitative comparison of the two couplings based on the average energy balance, which characterizes the collective behavior of the coupled neurons.

With instantaneous electrical coupling ($\tau=0$), a fully synchronized state (bursts and spikes within a burst are synchronized) is characterized by an average energy balance of zero. This fact is a consequence of the nature of the coupling term, which tends to zero when the two neurons are fully synchronized. This means that the average energy absorbed by the membrane is perfectly balanced by the average energy output (see Figure \ref{figure3}.

However, for instantaneous chemical coupling, the fully synchronized state cannot be reached for any value of synaptic coupling strength $g_c$, and further increasing $g_c$ only leads the coupled neurons to reach a steady state. This state is characterized by a positive average energy balance maintained mainly by the contribution of the synapse (see Figure \ref{figure4}).

In the presence of time delays, we find that coupled neurons can reach fully synchronized states for both electrical and chemical coupling at very low values of coupling strengths. These optimal settings of time delay and coupling strength correspond to the minima of the average energy variation (see Figure \ref{figure6}).

\newpage
\bibliographystyle{elsarticle-num-names}
\bibliography{references}

\end{document}